\documentstyle[twoside,fleqn,espcrc2,psfig]{article}

\newcommand{\AmS}{{\protect\the\textfont2
  A\kern-.1667em\lower.5ex\hbox{M}\kern-.125emS}}

\newcommand{\be}{\begin{equation}}
\newcommand{\ee}{\end{equation}}
\newcommand{\ben}{\begin{eqnarray}}
\newcommand{\een}{\end{eqnarray}}

\title{Quenched charmonium near the continuum limit
\thanks{Talk presented by M.~Garc\'{\i}a~P\'erez}}

\author{QCD-TARO Collaboration:
        S.~Choe${}^a$,
        Ph.~de~Forcrand${}^b$,
        M.~Garc\'{\i}a~P\'erez${}^c$,
        S.~ Hioki${}^d$,
        Y.~Liu$^{\rm a}$,
        H.~Matsufuru${}^e$,
        O.~Miyamura$^{\rm a}$,
        A.~Nakamura${}^f$,
        I.-O.~Stamatescu${}^{g,h}$,
        T.~Takaishi${}^i$,
        and
        T.~Umeda${}^j$\\
\vspace{4mm}
${}^a$ Department of Physics, Hiroshima University,
       Higashi-Hiroshima 739-8526, Japan\\
${}^b$ Institut f\"ur Theoretische Physik,
       ETH-H\"onggerberg, CH-8093 Z\"urich, Switzerland \\
${}^c$ Theory Division, CERN, CH-1211 Geneva 23, Switzerland \\
${}^d$ Department of Physics, Tezukayama University,
       Nara 631-8501, Japan \\
${}^e$ Yukawa Institute for Theoretical Physics,
       Kyoto University,  Kyoto 606-8502, Japan \\
${}^f$ IMC, Hiroshima University, Higashi-Hiroshima 739-8521, Japan \\
${}^g$ Institut f\"ur Theoretische Physik, Universit\"at
       Heidelberg,  D-69120 Heidelberg, Germany \\
${}^h$ FEST, Schmeilweg 5, D-69118 Heidelberg, Germany \\
${}^i$ Hiroshima University of Economics, Hiroshima 731-0192, Japan \\
${}^j$ Center for Computational Physics, University of Tsukuba,
       Tsukuba 305-8577, Japan \\
}

\begin{document}

\begin{abstract}
 We study relativistic charmonium on very fine quenched
 lattices ($\beta=6.4$ and 6.6). We concentrate on the calculation of 
 the hyperfine splitting between $\eta_c$ and $J/\psi$, aiming for a controlled
 continuum extrapolation of this quantity. Results for the $\eta_c$ and $J/\psi$ 
 wave functions are also presented.
\end{abstract}
\maketitle

\section{Introduction}
Charmonium spectroscopy on the lattice is not at all straightforward.
Charm is too heavy for most current simulations -- typically $m a > 1$ --
but too light to blindly rely on the heavy quark approximation --
for $c \bar{c}$, $v^2/c^2 \sim 0.3$. A good probe of relativistic effects
is the hyperfine splitting between the $^3 S_1$ and the $^1 S_0$ states, which 
for charmonium is $\Delta M\!=\!117$ MeV. Lattice quenched calculations 
underestimate 
$\Delta M$ by about $30-50\%$  \cite{Cppacs,UKQCD}, the prediction 
from NRQCD being $\Delta M\!= \!55(5)$ MeV \cite{Trottier}. The discrepancy 
could be due to the quenched approximation. However,  first 
estimates including dynamical quarks seem to indicate that they  account 
for only $\sim 10\%$ of the difference \cite{Din}.
Here we compute $\Delta M$ on fine lattices within the relativistic formalism, 
aiming for the extraction of a controlled quenched continuum limit.

Two approximations will be made: $(i)$ quenched, $(ii)$ OZI, meaning that 
Zweig-rule forbidden diagrams, although contributing to singlet mesons like
charmonium, will not be included.

\section{Results}

Configurations are generated with the standard Wilson action
on a $32^3 \!\times\! 96$ lattice at $\beta=6.4$ and 6.6. 
Quark propagators are computed using Wilson, tree-level clover and 
tadpole improved clover Dirac operators (at $\beta=6.6$, 
$c_{sw}^{\rm tadp}=1.388$ while the non-perturbative  $c_{sw}=1.467$ 
\cite{Alpha}). We have 60 configurations at each $\beta$  value.

\subsection{Hyperfine splitting}
\begin{figure}[htb]
\vspace{2.6cm}
\includegraphics{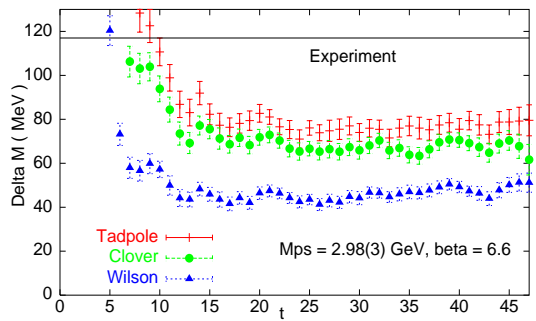}
\caption{Hyperfine splitting effective mass.}
\label{fig1}
\end{figure}

$\Delta M$ is computed from the ratio of vector to pseudoscalar correlators.
Fig. \ref{fig1} shows the effective mass at $\beta=6.6$
for a pseudoscalar mass tuned close to the physical $\eta_c$ mass 
(the scale set by $r_0$ from  \cite{Sommer}).
Even on this fine lattice, the sensitivity of $\Delta M$ 
to the Dirac operator is large, tadpole improvement giving
the best estimate.  The total set of results for the 
pseudoscalar mass and the hyperfine splitting is presented in 
Table \ref{table1}.
\vspace{-0.6cm}

\begin{table}[htb]
\caption{Pseudoscalar mass ($M_{\rm ps}$) and hyperfine splitting ($\Delta M$)
in lattice units. W, Cl and Tadp denote Wilson, clover and tadpole
improved clover operators.}
\label{table1} 
\newcommand{\m}{\hphantom{$-$}}
\newcommand{\cc}[1]{\multicolumn{1}{c}{#1}}
\begin{tabular}{@{}lllll}
\hline
Op.&\cc{$\beta$} & \cc{$\kappa$} & \cc{$a M_{\rm ps}$} &\cc{$a \Delta M$} \\
\hline
W   &     6.4   & 0.1389     & 0.7632(8)      & 0.01008(30) \\
Cl  &     6.4   & 0.1324     & 0.7645(8)      & 0.01664(44) \\
W   &     6.6   & 0.1415     & 0.5228(9)      & 0.01176(23) \\
W   &     6.6   & 0.1400     & 0.5993(9)      & 0.00915(24) \\
W   &     6.6   & 0.1385     & 0.6729(9)      & 0.00763(19) \\
W   &     6.6   & 0.1375     & 0.7205(9)      & 0.00665(17) \\
Cl  &     6.6   & 0.13225    & 0.6735(8)      & 0.01157(35) \\
Cl  &     6.6   & 0.1335     & 0.5969(9)      & 0.01357(34) \\
Tadp&     6.6   & 0.1158     & 0.6700(9)      & 0.01371(34) \\
Tadp&     6.6   & 0.1167     & 0.6031(9)      & 0.01538(38) \\
\hline
\end{tabular}\\[2pt]
\end{table}
\vspace{-0.6cm}

\begin{figure}[htb]
\vspace{3.6cm}
\includegraphics{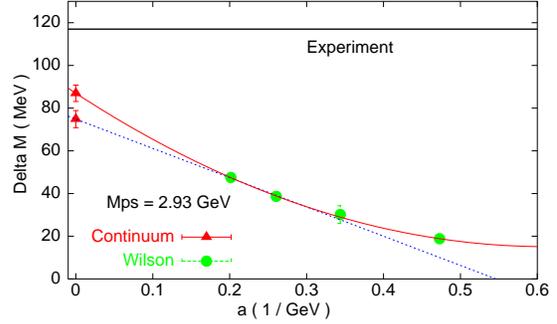}
\caption{Continuum extrapolation of $\Delta M$ with the Wilson Dirac operator.}
\label{fig2}
\end{figure}

\begin{figure}[htb]
\vspace{2.7cm}
\includegraphics{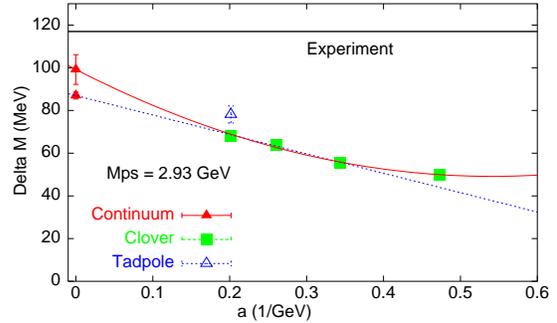}
\caption{Continuum extrapolation of $\Delta M$ with the tree-level clover 
improved Dirac operator.}
\label{fig3}
\vspace{-0.5cm}
\end{figure}

To perform the continuum extrapolation we use $\beta\!=\!6.0$ and 6.2
data from UKQCD \cite{UKQCD,Collins}, available for
both Wilson and tree-level clover Dirac operators. Fitting the 
dependence of $\Delta M$ versus $1/M_{\rm ps}$ at fixed $\beta$ 
allows to interpolate $\Delta M(\beta)$ to a common value of 
the pseudoscalar mass: $M_{\rm ps}\!=\!2.93$ GeV, which is our 
value at $\beta\!=\!6.4$ (experimentally $M_{\eta_c}\!=\!2.98$ GeV). 
In Figures \ref{fig2}, \ref{fig3} we present the continuum extrapolation of 
the Wilson and tree-level clover hyperfine splittings, respectively. 
Note that: 
$(i)$ A linear extrapolation in $a$ from $\beta\!=\!6.0$ is not 
justified. This is a clear indication of strong 
lattice artifacts on this coarse lattice 
($a M_{\rm ps} \approx 1.5$). $(ii)$ A continuum extrapolation is mandatory: 
at all the simulated $\beta$ values the
difference between Wilson and clover data is large but their 
extrapolations are consistent with each other. We quote, from the quadratic (linear) fits,
$\Delta M \!=\! 87\pm 4 \ (75\pm 4)$ MeV and $\Delta M\! = \!99\pm 7 \ (87\pm 2)$ MeV
for Wilson and clover respectively.
We have also included in Fig. \ref{fig3} the value 
of $\Delta M$ obtained from tadpole improvement at $\beta=6.6$. A 
more extensive calculation including  non-perturbative
improved clover data is in progress.


\begin{table*}[htb]
\caption{Hyperfine splitting from different approaches.
The scale is set by $r_0$ or the spin averaged $^1P_1\!-\! 1\overline{S}$
splitting.}
\label{table.all}
\newcommand{\m}{\hphantom{$-$}}
\newcommand{\cc}[1]{\multicolumn{1}{c}{#1}}
\renewcommand{\tabcolsep}{0.73pc} 
\renewcommand{\arraystretch}{1.07} 
\begin{tabular}{@{}llllll}
\hline
&This work & This work &CP-PACS \cite{Cppacs} &
CP-PACS \cite{Cppacs} &latest NRQCD\cite{Trottier} \\
&Relativistic & Relativistic &Relativistic & Relativistic &Non relativistic \\
&Isotropic & Isotropic &Anisotropic &Anisotropic & \\
&Continuum & Tadpole $\beta=6.6$&Continuum & Continuum & \\
\hline
\vspace{0.07cm}
Scale& $\ \ \ r_0$ &  $\ r_0$ & $\ r_0$ &
 $^1P_1- 1\overline{S}$ & $^1P_1- 1\overline{S}$ \\
$\Delta M$(MeV)&  83-106  & 78(2) & 65(1) & \m82(2)& \m55(5) \\
\hline
\end{tabular}\\[2pt]
\vspace{-0.46cm}
\end{table*}

\subsection{Wave functions}

Within the naive non-relativistic approximation it is easy 
to understand the origin of the hyperfine splitting.
This approximation amounts to solving the Schr\"odinger equation 
in a non-relativistic potential and dealing with relativistic corrections 
in perturbation theory. To zeroth order $^3 S_1$ and $^1 S_0$ states
are degenerate. The first order correction to the mass is the
average of the spin-spin interaction giving
$\Delta M \propto \alpha_s(m_q) |\Psi(0)|^2/m_q^2 $, with $\Psi(0)$
the value of the non-relativistic wave function at the origin.
Perturbative corrections to the wave function also depend on
the spin; to lowest order, $|\Psi(0)|$ increases (decreases)
for the pseudoscalar (vector). 

\begin{figure}[htb]
\vspace{7.0cm}
\includegraphics{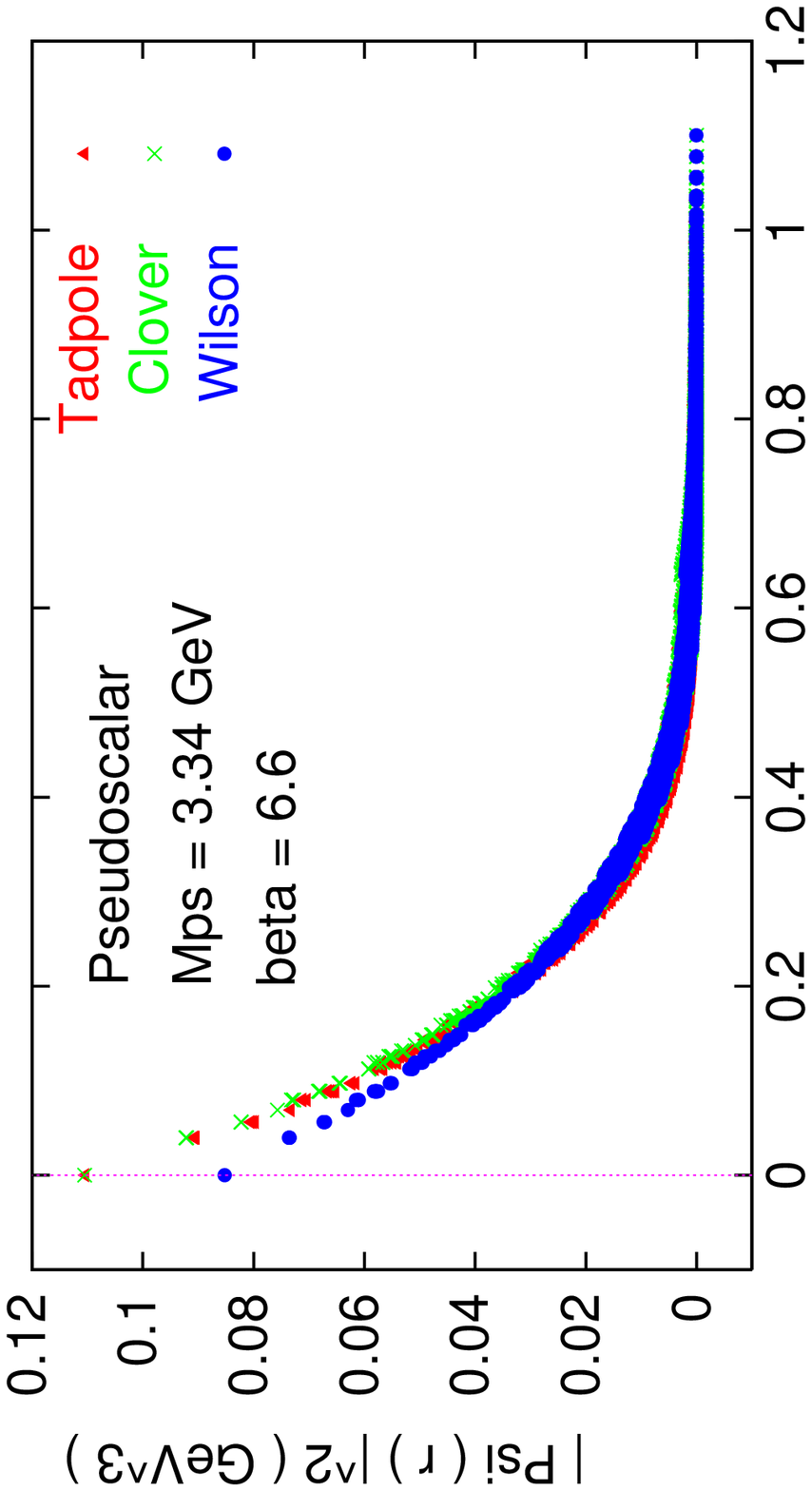}
\includegraphics{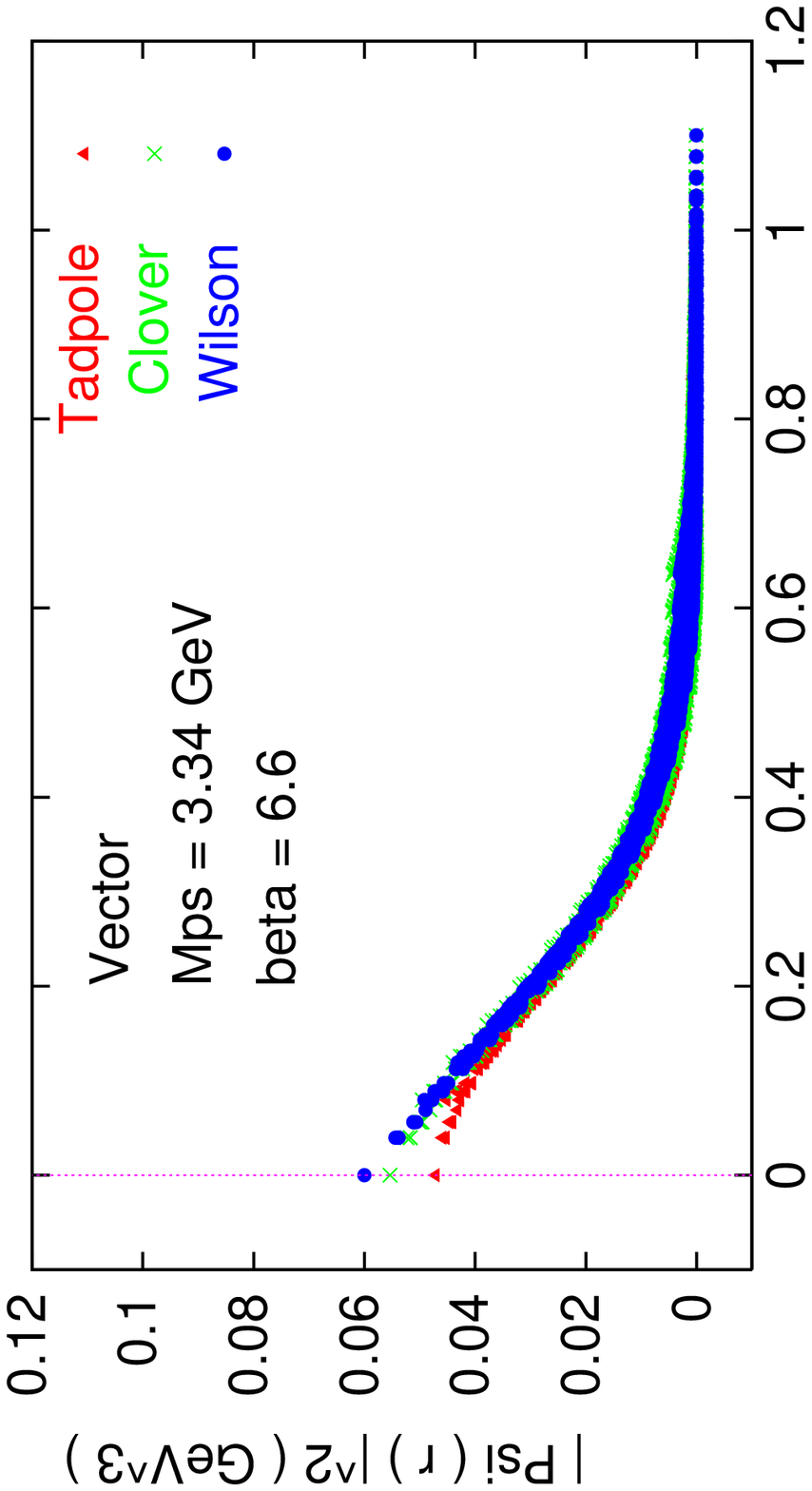}
\caption{Pseudoscalar (top) and vector (bottom) wave functions.}
\label{fig.wavef}
\end{figure}

We have extracted gauge invariant wave functions from lattice 4-point 
functions: $ \Psi (x)= \langle c \bar{c} | (\psi_c^\dagger \psi_c)(0) 
(\psi_{\bar{c}}^\dagger \psi_{\bar{c}})(x) | c \bar{c} \rangle$ . 
A comparison between pseudoscalar and vector wave functions
is presented in Fig. \ref{fig.wavef}. The observed pattern corroborates
qualitatively the predictions of the heavy-quark model. We also exhibit 
the difference between the wave functions extracted from Wilson, clover 
and tadpole improved clover Dirac operators. As already observed for the 
hyperfine splitting, relativistic effects (implying degeneracy breaking 
between $^3 S_1$ and $^1 S_0$) increase as we improve the 
fermionic action. 

\vspace{-1.0cm}
\subsection{Conclusions}

Table \ref{table.all} compiles data from 
\cite{Cppacs,Trottier}. We compare with an interval
covering our Wilson and clover continuum extrapolations.
Using in all cases $r_0$ to fix the scale, our 
relativistic extrapolation gets much closer to the experimental value. 
Note that this is also the case
for our $\beta\!=\!6.6$ tadpole improved clover data, which gives
$\Delta M\!=\!78(2)$ MeV. 

We conclude that a careful and controlled quenched continuum extrapolation
is now feasible, and gets closer to the experimental value than
previous determinations.  Dynamical quark effects, which
presumably account for the bulk of the remaining disagreement with experiment,
appear to be O(10-20)$\%$, which is consistent with the change in the
coupling constant $\alpha(N_f\!=\!0) \!\rightarrow\! \alpha(N_f\!=\!2)$ 
entering in the non-relativistic expression. OZI suppressed contributions
are expected to be small for these heavy quark masses, but we cannot rule out the 
possibility that they give a non negligible contribution to $\Delta M$, 
which could be enhanced if there is mixing with glueballs. Work is in
progress to improve the extrapolations and to extract masses for other channels.
\vspace{-0.15cm}
\subsection*{Acknowledgements}

We thank Sara Collins for providing the raw UKQCD data and 
Stephan Sint for pointing out the possible relevance of OZI 
suppressed diagrams.

\end{document}